\documentclass[12pt,oneside]{article}

\begin{document}

\title{Understanding the landscape\thanks{
Based on comments made at the 23rd Solvay Conference,
December 2005, Brussels}}

\author{Michael R. Douglas\\
NHETC, Rutgers University\\
Piscataway, NJ 08855--0849, USA\\
{\it and} 
I.H.E.S., Le Bois-Marie, Bures-sur-Yvette, 91440 France\\
{\tt mrd@physics.rutgers.edu}
}

\maketitle

\setcounter{equation}{0}
\newfam\black
\font\blackboard=msbm10 
\font\blackboards=msbm7
\font\blackboardss=msbm5
\textfont\black=\blackboard
\scriptfont\black=\blackboards
\scriptscriptfont\black=\blackboardss
\def\Bbb#1{{\fam\black\relax#1}}
\def\scrl{{\ell}}
\def\sgn{{\rm sgn\ }}
\def\etal{{\it et.al.}}
\def\slashslash{{/\hskip-0.2em/}}
\def\CA{{\cal A}}
\def\CC{{\cal C}}
\def\CD{{\cal D}}
\def\CF{{\cal F}}
\def\CH{{\cal H}}
\def\CI{{\cal I}}
\def\CJ{{\cal J}}
\def\CL{{\cal L}}
\def\CM{{\cal M}}
\def\CN{{\cal N}}
\def\CO{{\cal O}}
\def\CP{{\cal P}}
\def\CT{{\cal T}}
\def\CW{{\cal W}}
\def\CZ{{\cal Z}}
\def\BC{\Bbb{C}}
\def\BH{\Bbb{H}}
\def\BM{\Bbb{M}}
\def\BP{\Bbb{P}}
\def\BR{\Bbb{R}}
\def\BX{\Bbb{X}}
\def\BZ{\Bbb{Z}}
\def\mapr{\mathop{\longrightarrow}\limits}
\def\half{{1\over 2}}
\def\eV{~{\rm eV}}
\def\GeV{~{\rm GeV}}
\def\TeV{~{\rm TeV}}
\def\Coh{{\rm Coh}~}
\def\Cohc{{\rm Coh}_c~}
\def\Mod{{\rm Mod}~}
\def\ind{{\rm ind}~}
\def\Tr{{\rm Tr}~}
\def\tr{{\rm tr}~}
\def\grad{{\rm grad}~}
\def\CY#1{CY$_#1$}
\def\rk{{\rm rk}~}
\def\Re{{\rm Re}~}
\def\Im{{\rm Im}~}
\def\Hom{{\rm Hom}}
\def\Ext{{\rm Ext}}
\def\Volume{{\rm Volume~}}
\def\Vol{{\rm Vol~}}
\def\Stab{{\rm Stab}}
\def\End{{\rm End~}}
\def\ib{{\bar i}}
\def\Dbar{{\bar D}}
\def\jb{{\bar j}}
\def\ab{{\bar a}}
\def\db{{\bar d}}
\def\zb{{\bar z}}
\def\pp{\partial}
\def\pb{{\bar\partial}}
\def\I{{I}}
\def\II{{II}}
\def\IIa{{IIa}}
\def\IIb{{IIb}}
\def\vev#1{{\langle#1\rangle}}
\def\vvev#1{{\langle\langle#1\rangle\rangle}}
\def\ket#1{{|#1\rangle}}
\def\Dslash{\rlap{\hskip0.2em/}D}
\def\dual{{v}} 
\def\intersect{\cdot}
\def\Nbar{{\bar N}}
\def\NRR{{N_{RR}}}
\def\NNS{{N_{NS}}}
\def\bigvev#1{\bigg\langle{#1}\bigg\rangle}
\def\boldclass{\bf\sf}
\def\P{{\boldclass P}}
\def\NP{{\boldclass NP}}

\section{Historical analogies}

At a meeting such as this, one is tempted to compare our present
struggles to understand string theory, and to find clearer evidence
for or against the claim that it describes our universe, with the deep
issues discussed at past Solvay conferences.  Now my experience has been
that when visiting Belgium, giving in to temptation is generally
a good thing to do, so with no further apology let us proceed.

As was beautifully described here by Peter Galison, the 1911 meeting
focused on the theory of radiation, and the quantum hypotheses
invented to explain black body radiation and the photoelectric effect.
These were simple descriptions of simple phenomena, which suggested a
new paradigm.
This was to accept the basic structure of 
previous models, but 
modify the laws of classical mechanics by inventing new,
somewhat {\it ad hoc} rules governing quantum phenomena.
This paradigm soon scored a great success in Bohr's theory of the
hydrogen atom.  The discovery of the electron and Rutherford's
scattering experiments had suggested modeling
an atom as analogous to a planetary system.  But while planetary
configurations are described by continuous parameters, real atoms have
a unique ground state, well-defined spectral lines
associated with transitions from excited states, etc.  
From Bohr's postulate that the action of
an allowed trajectory was quantized, he was able to deduce all of
these features and make precise numerical predictions.

While very successful, it was soon found that this did not work for
more complicated atoms like helium.  A true quantum mechanics had to
be developed.  Most of its essential ideas had appeared by the 1927
meeting.  Although the intuitions behind the Bohr atom turned out to
be correct, making them precise required existing but
unfamiliar mathematics, such as the theories of infinite
dimensional matrices, and wave equations in configuration space.

Are there fruitful analogies between these long-ago problems and our
own?  What is the key issue we should discuss in 2005?  What are our
hydrogen atom(s)?

If we have them, they are clearly the maximally
supersymmetric theories, whose basic physics was elucidated in the
second superstring revolution of 1994--98.  It's too bad we can't 
use them to describe
real world physics.  But they have precise and pretty
formulations, and can be used to model one system we believe exists in
our universe, the near-extremal black hole.  We now have
microscopic models of black holes, which explain their entropy.

Perhaps we can place our position as analogous to the period between 1913
and 1927.\footnote{
A similar analogy was made by David Gross in talks given around 2000.
However, to judge from his talk here, he now has serious reservations
about it.
}
Starting from our simple and attractive maximally supersymmetric theories,
we are now combining their ingredients in a somewhat {\it ad hoc} way,
to construct $N=1$ and nonsupersymmetric
theories, loose analogs of helium, molecules, and more complicated
systems.  The Standard Model, with its $19$ parameters, has a
complexity perhaps comparable to a large atom or small molecule.  
The difficulty of our
present struggles to
reproduce its observed intricacies and the underlying infrastructure
(moduli stabilization, supersymmetry breaking), 
discussed here by Kallosh, L\"ust and others, are probably a
sign that we have not yet found the best mathematical framework.

\section{The chemical analogy}

What might this ``best mathematical framework'' be?
And would knowing it help with the central problems preventing
us from making definite predictions and testing the theory?

In my opinion, the most serious obstacle to testing the theory is the
problem of vacuum multiplicity.  This has become acute with the recent
study of the string/M theory landscape.  We have a good reason to
think the theory has more than $10^{122}$ vacua, the
Weinberg-Banks-Abbott-Brown-Teitelboim-Bousso-Polchinski {\it et al}
solution to the cosmological constant problem.  Present computations
give estimates more like $10^{500}$ vacua.  We do not even know the
number of candidate vacua is finite.  Even granting that it is, the
problem of searching through all of them is daunting.  Perhaps 
{\it a priori} selection principles or measure factors will help, but
there is little agreement on what these might be.
We should furthermore admit that the explicit constructions of vacua and
other arguments supporting this picture, while improving, are not yet
incontrovertible.

We will shortly survey a few mathematical frameworks which may be
useful in coming to grips with the landscape, either directly or by
analogy.  They are generally not familiar to physicists.  I think the
main reason for this is that analogous problems in the past were
attacked in different, non-mathematical ways.  Let us expand a bit on
this point.

String theory is by no means the first example of an underlying simple
and unique framework describing a huge, difficult to comprehend
multiplicity of distinct solutions.  There is another one, very well known,
which we might consider as a source of analogies.

As condensed matter physicists never tire of reminding us, all of the
physical properties of matter in the everyday world, and the diversity
of chemistry, follow in principle from a well established ``theory of
everything,'' the Schr\"odinger equations governing a collection of
electrons and nuclei.  Learning even the rough outlines of the
classification of its solutions takes years and forms the core of
entire academic disciplines: chemistry, material science, and their
various interdisciplinary and applied relatives.

Of course, most of this knowledge was first discovered empirically, by
finding, creating and analyzing different substances, with the
theoretical framework coming much later.
But suppose we were given the Schr\"odinger equation and Coulomb
potential without this body of empirical knowledge?  Discovering
the basics of chemistry would be a formidable project, and 
there are many more layers of structure to elucidate before one would 
reach the phenomena usually discussed in condensed matter physics:
phase transitions, strong correlations, topological structures and
defects, and so on.

As in my talk at String 2003, one can develop this analogy, by
imagining beings who are embedded in an effectively infinite crystal,
and can only do low energy experiments.  Say they can observe the
low-lying phonon spectrum, measure low frequency conductivity, and so
on.  Suppose among their experiments they can create electron-hole
bound states, and based on phenomenological models of these they
hypothesize the Schr\"odinger equation.  They would have some
empirical information, but not the ability to manipulate atoms and
create new molecules.  How long would it take them to come up with the
idea of crystal lattices of molecules, and how much longer would it
take them to identify the one which matched their data?

Now, consider the impressive body of knowledge string theorists
developed in the late 1990's, assembling quasi-realistic
compactifications out of local constituents such as branes,
singularities, and so on.  Individual constituents are simple, their
basic properties largely determined by the representation theory of
the maximal supersymmetry algebras in various dimensions.  The rules
for combining pairs of objects, such as intersecting branes or branes
wrapping cycles -- which combinations preserve supersymmetry, and what
light states appear -- are not complicated either.  What is
complicated is the combination of the whole required to duplicate the
Standard Model, stabilize moduli, break supersymmetry and the rest.
Perhaps all this is more analogous to chemistry than we would like to
admit.

Other parallels can be drawn.  For example,\footnote{
As recalled here by Joe Polchinski.}
according to standard nuclear physics, the lowest
energy state of a collection of electrons, protons and neutrons is a
collection of ${}^{62}{\rm Ni}$ atoms, and thus almost all molecules in
the real world are unstable under nuclear processes.  Suppose this were
the case for our crystal dwellers as well.  After learning
about these processes, 
they might come to a deep paradox: how can
atoms other than nickel
exist at all?  Of course, because of Coulomb barriers, the
lifetime of matter is exceedingly long, but still finite, just as is
claimed for the metastable de Sitter vacua of KKLT.

Perhaps all this is a nightmare from which we will awake, the history
of Kekul\'e's dream being repeated as farce.  If so, all our previous
experience as physicists suggests that the key to the problem will be
to identify some sort of {\bf simplicity} which we have not seen in
the problem so far.  One might look for it in the physics of some dual
or emergent formulation.  But one might also look for it in
mathematics.  It is not crazy to suppose that the only consistent vacua are
those which respect some principle or have some property which would
only be apparent in an exact treatment.  But what is that exact treatment
going to look like?  The ones we have now cannot be formulated without
bringing in mathematics such as the geometry of Calabi-Yau manifolds,
or the category theory underlying topological string theory.  If we 
ever find exact descriptions of $N=1$ or broken supersymmetry
vacua, surely this will be by uncovering even more subtle mathematical
structures.

But suppose the landscape in its present shape is real, and the key to
the problem is to manage and abstract something useful out of its {\bf
complexity}.  The tools we will need may not be those we traditionally
associated with fundamental physics, but might be inspired by other
parts of physics and even other disciplines.  But such inspiration can
not be too direct; the actual problems are too different.  Again, we
are probably better off looking to mathematical developments which
capture the essence of the ideas and then generalize them, as more
likely to be relevant.

On further developing these analogies, one realizes that we do not
know even the most basic organizing principles of the stringy
landscape.  For the landscape of chemistry, these are the existence of
atoms, the maximal atomic number, and the
facts that each atom (independent of its type) takes up a
roughly equal volume in three-dimensional space and that binding
interactions are local.  These already determine the general features
of matter, such as the fact that densities of solids range from
$1$--$20~{\rm g/cm^3}$.  Conjectures on the finite number of string
vacua, on bounds on the number of massless fields or ranks of gauge
groups, and so on, are suggestions for analogous general features of
string vacua.  But even knowing these, we would want organizing principles.
The following brief overviews should be read with this question in mind.

\section{Two-dimensional CFT}

This is not everything, but a large swathe through the landscape.
We do not understand it well enough.
In particular, the often used concept of ``the space of 2d CFT's,'' 
of obvious relevance for our questions, has
never been given any precise meaning.  

A prototype might be found in the mathematical theory of the
space of all Riemannian manifolds.  This exists and is useful for
broad general statements.  We recall Cheeger's theorem \cite{cheeger}:

A set of manifolds with metrics
$\{X_i\}$, satisfying the following bounds,
\begin{enumerate}
\item ${\rm diameter}(X_i) < d_{max}$
\item $\Volume(X_i) > V_{min}$
\item Curvature $K$ satisfies $|K(X_i)| < K_{max}$ at every point,
\end{enumerate}
contains a finite number of distinct homeomorphism types
(and diffeomorphism types in $D\ne 4$).

Since (2) and (3) are conditions for validity of supergravity, while
(1) with $d_{max} \sim 10 \mu m$
follows from the validity 
of the gravitational inverse square law down to this distance,
this theorem implies that there are finitely many manifolds which
can be used for candidate supergravity compactifications
\cite{str05,ach}

This and similar theorems are based on more general quasi-topological
statements such as Cheeger-Gromov precompactness of the space of metrics --
{\it i.e.}, infinite sequences have Cauchy subsequences, and cannot
``run off to infinity.''  This is shown by constructions which break
any manifold down into a finite number of coordinate patches, and showing
that these patches and their gluing can be described by a finite amount
of data.

Could we make any statement like this for the space of CFT's? (a question
raised by Kontsevich).  The diameter bound becomes a lower bound 
$\Delta_{min}$ on the
operator dimensions (eigenvalues of $L_0+\bar L_0$).  We also need to fix $c$.
Then, the question seems well posed, but we have no clear approach to it.
Copying the approach in terms of coordinate patches does not seem right.

The key point in defining any ``space'' of anything is to put a 
topology on the set of objects.  Something less abstract from which 
a topology
can be derived is a distance between pairs of objects $d(X,Y)$ 
which satisfies the axioms of a metric, so that it can be used to define
neighborhoods.

The usual operator approach to CFT, with a Hilbert space $\cal H$,
the Virasoro algebras with $H=L_0+\bar L_0$, and the operator product
algebra, is very analogous to spectral geometry:
\begin{center}
$L_0+\bar L_0$ eigenvalues $\sim$ spectrum of Laplacian 
$\Delta$ \\
o.p.e. algebra $\sim$ algebra of functions on a manifold
\end{center}
Of course the o.p.e. algebra is not a 
standard commutative algebra and this is analogy,
but a fairly close one.

A definition of a distance between a pair of manifolds with metric,
based on spectral geometry, is given in B\'erard, Besson, and
Gallot \cite{BBG}.  The idea is to consider
the entire list of eigenfunctions $\psi_i(x)$ of the Laplacian,
$$
\Delta \psi_i = \lambda_i \psi_i ,
$$
as defining an embedding $\Psi$ of the manifold into ${\scrl}_2$, the
Hilbert space of semi-infinite sequences (indexed by $i$):
$$
\Psi: x \rightarrow \{e^{-t\lambda_1}\psi_1(x),
e^{-t\lambda_2}\psi_2(x),\ldots,e^{-t\lambda_n}\psi_n(x),\ldots\} .
$$
We weigh by $e^{-t\lambda_i}$ for some fixed $t$
to get convergence in 
$\scrl_2$.

Then, the distance between two manifolds $M$ and $M'$ is the Hausdorff
distance $d$ between their embeddings in ${\scrl}_2$.  Roughly, this is the
amount $\Psi(M)$ has to be ``fuzzed out'' to cover $\Psi(M')$.

In principle this definition might be directly adapted to CFT, where the
$x$ label boundary states $\ket{x}$
(which are the analog of points) and the $\psi_i(x)$ are 
their overlaps with closed string states $\ket{\phi_i}$,
$$
x \rightarrow \vev{\phi_i|e^{-t(L_0+\bar L_0)}|x} .
$$

Another candidate definition would use the o.p.e. coefficients
$$
\phi_i \phi_j \rightarrow \sum C_{ij}^k 
(z_i-z_j)^{\Delta_k-\Delta_i-\Delta_j}\ \phi_k
$$
for all operators with dimensions between $\Delta_{min}$ and some
$\Delta_{max}$ (one needs to show that this choice
drops out), again weighted by
$e^{-t(L_0+\bar L_0)}$.  The distance between a pair of CFT's would then
be the $\ell_2$ norm of the differences between these sets of numbers.

While abstract, this would make precise the idea of the ``space of all
2D CFT's'' and give a foundation for mapping it out.

\section{Topological open strings and derived categories}

This gives an example in which we actually know ``the space of all X''
in string theory.  It is based on the discussion of boundary
conditions and operators in CFT, which satisfy an operator 
product algebra with
the usual non-commutativity of open strings.  If
we modify the theory to obtain a subset of dimension zero operators
(by twisting to get a topological open string, 
taking the Seiberg-Witten limit in a $B$ field, etc.), the o.p.e.
becomes a standard associative but non-commutative
algebra.  This brings us into the realm of noncommutative geometry.

There are many types of noncommutative geometry.  For the standard
topological string obtained by twisting an $N=2$ theory,
the most appropriate
is based on algebraic geometry.  As described at the Van den Bergh
2004 Francqui prize colloquium, this is a highly developed subject,
which forms the backdrop to quiver gauge theories, D-branes on 
Calabi-Yau manifolds, and so on.  

One can summarize the theory of D-branes on a Calabi-Yau $X$ in these
terms as the ``Pi-stable objects in the derived category $D(\Coh
X)$,'' as reviewed in \cite{Dbranes}.  Although abstract, the
underlying idea is simple and physical.  It is that all branes can be
understood as bound states of a finite list of ``generating branes,''
one for each generator of K theory, and their antibranes.  The bound
states are produced by tachyon condensation.  Varying the Calabi-Yau
moduli can vary masses of these condensing fields, and if one goes
from tachyonic to massive, a bound state becomes unstable.

This leads to a description of all D-branes, and ``geometric''
pictures for all the processes of topology change which were
considered ``non-geometric'' from the purely closed string point of
view.  For example, in a flop transition, an $S^2$ $\Sigma$ is cut out
and replaced with another $S^2$ $\Sigma'$ in a topologically different
embedding.  In the derived category picture, what happens is that the
brane wrapped on $\Sigma$, and all D$0$'s (points) on $\Sigma$, go
unstable at the flop transition, to be replaced by new branes on
$\Sigma'$.

The general idea of combining classical geometric objects, using stringy
rules of combination, and then extrapolating to get a more general type
of geometry, should be widely useful.

\section{Computational complexity theory}

How hard is the problem of finding quasi-realistic string vacua?
Computer scientists classify problems
of varying degrees of difficulty:
\begin{itemize}
\item {\P} can be solved in time polynomial in the size of the input.
\item
An {\NP} problem has a solution which can be checked in 
polynomial time, but is far harder to find, typically 
requiring a search through all candidate solutions.
\item
An {\NP}-complete problem is as hard as any {\NP} problem -- if any
of these can be solved quickly, they all can.
\end{itemize}

It turns out that many of the problems arising in the search for
string vacua are in {\NP} or even \NP-complete. \cite{ddone}  
For example, to
find the vacua in the Bousso-Polchinski model with cosmological
constant $10^{-122} M_{Planck}^4$, one may need to search through
$10^{122}$ candidates.

How did the universe do this?  We usually say that the ``multiverse''
did it -- many were tried, and we live in one that succeeded.
But some problems are too difficult for the multiverse to solve in
polynomial time.  This is made precise by Aaronson's definition
of an ``anthropic computer.'' \cite{Aaronson}

Using these ideas, Denef and I \cite{ddtwo} have argued that the vacuum
selected by the measure factor $\exp 1/\Lambda$ cannot be
found by a quantum computer, working in polynomial time, even with
anthropic postselection.  Thus, if a cosmological model realizes this
measure factor (and many other preselection principles which can be
expressed as optimizing a function), it is doing something more
powerful than such a computer.

Some cosmological models (e.g. eternal inflation) explicitly postulate
exponentially long times, or other violations of our hypotheses.  But
for other possible theories, for example a field theory dual to eternal
inflation, this might lead to a paradox.

\section{Conclusions}

We believe string theory has a set of solutions, some of which might
describe our world.  Even leaving aside the question of few vacua or
many, and organizing principles, perhaps the most basic
question about the landscape is whether it will turn out to be
more like mathematics, or more like chemistry.

Mathematical analogy: like classification of Lie groups, finite simple
groups, Calabi-Yau manifolds, etc.  Characterized by simple axioms
and huge symmetry groups.  In this vision, the overall
structure is simple, while the intricacies of our particular vacuum
originate in symmetry breaking analogous to that of more familiar
physical systems.

Chemical analogy: simple building blocks (atoms; here branes
and extended susy gauge theory sectors) largely determined
by symmetry.  However, these are combined in intricate ways which defy simple
characterization and require much study to master.

The current picture, as described here by Kallosh and L\"ust, 
seems more like chemistry.  Chemistry is a great science, after
all the industrial chemistry of soda is what made these 
wonderful conferences possible.  But it will
surely be a long time (if ever) before we can manipulate the
underlying constituents of our vacuum and produce new solutions, so 
this outcome would be less satisfying.  

Still, our role as physicists is not to hope that one or the other
picture turns out to be more correct, but to find the evidence from
experiment and theory which will show us which if any of our present
ideas are correct.

\vspace{5mm}

\noindent {\bf Acknowledgments} I would like to thank the
organizers and the Solvay Institute for a memorable meeting, 
and B. Acharya, F. Denef, M. Kontsevich, S. Zelditch and many others
for collaboration and discussion of these ideas.
This research was supported
in part by DOE grant DE-FG02-96ER40959.

\end{document}